\documentclass[a4paper, 11pt, onecolumn]{article}

\usepackage[margin=2.0cm]{geometry}

\usepackage{calc}
\usepackage{enumitem}

\usepackage{booktabs}
\usepackage{array}

\usepackage[numbers, sort&compress]{natbib}
\usepackage{amsmath, amssymb, amsthm}

\usepackage[utf8]{inputenc}

\usepackage[american]{babel}
\usepackage{csquotes}

\usepackage{tikz,lipsum,lmodern}
\usepackage[most]{tcolorbox}

\usepackage[ttscale=.875]{libertine}
\usepackage[scaled=.96]{zi4}

\usepackage{nicefrac}

\usepackage{graphicx}

\usepackage[labelsep=period]{caption}

\usepackage{xcolor}
\definecolor{blendedblue}{rgb}{0.2, 0.2, 0.6}
\definecolor{blendedred}{rgb}{0.8, 0.2, 0.2}

\usepackage[bookmarks   = true,
            citecolor   = blendedblue,
            colorlinks  = true,
            linkcolor   = blendedblue,
            urlcolor    = blendedblue,
            citecolor   = blendedblue,
            linktocpage = false,
            hyperindex  = true,
            linkbordercolor = white]{hyperref}
\hypersetup{colorlinks = true}

\newcommand*{\figref}[2][]{%
  \hyperref[{fig:#2}]{%
    Fig.~\ref*{fig:#2}%
    \ifx\\#1\\%
    \else
      \,#1%
    \fi
  }%
}
\usepackage{doi}
\usepackage{url}

\usepackage{fancyhdr}
\makeatletter
\renewcommand{\maketitle}{\bgroup\setlength{\parindent}{0pt}
\begin{flushleft}
  \textbf{\huge\@title\\}
  \vspace{5mm}
  \@author
\end{flushleft}\egroup
}
\makeatother

\usepackage{tikz}
\usetikzlibrary{positioning,shapes,shadows,arrows}

\title{
  \fbox{
  \begin{minipage}{\textwidth-.9cm}
    \small{
      This is the peer reviewed version of the following article: Mahmood T., Wittenberg, P.,
      Zwetsloot, I.M., Wang, H., Tsui, K.L. (2019). Monitoring data quality for telehealth
      systems in the presence of missing data. International Journal of Medical Informatics
      126, 156--163, which has been published in final form at
      \href{https://dx.doi.org/10.1016/j.ijmedinf.2019.03.011}{https://dx.doi.org/10.1016/j.ijmedinf.2019.03.011}.
      This manuscript version is made available under the
      \href{https://creativecommons.org/licenses/by-nc-nd/4.0}{CC-BY-NC-ND 4.0} license.}
  \end{minipage}
} \vspace{2mm}\\
  \textbf{Monitoring data quality for telehealth systems in the presence of missing data}
}

\author{%
   \begin{large}
     \textbf{Tahir Mahmood,}$^{\textrm{a},1}$
     \href{https://orcid.org/0000-0002-8748-5949}{\includegraphics[width=8pt]{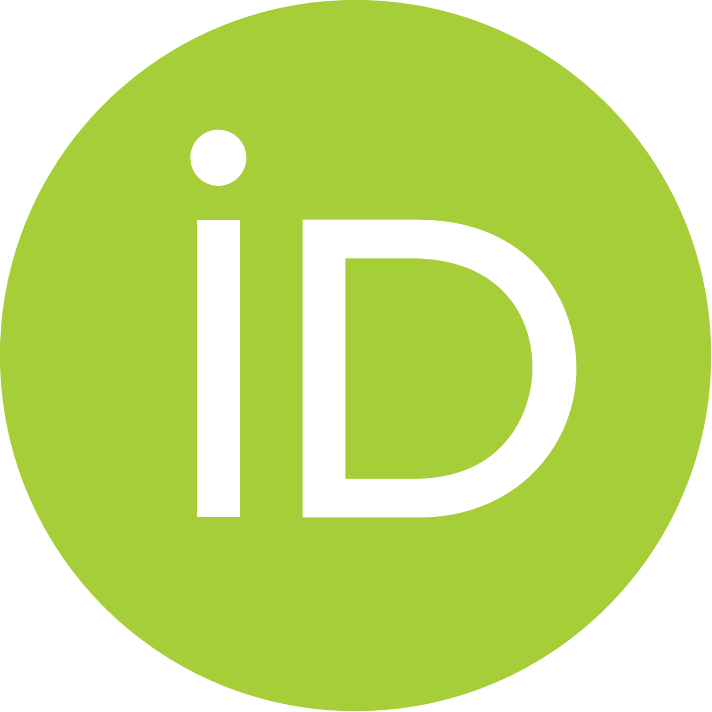}}
     \textbf{Philipp Wittenberg,}$^{\textrm{b},1}$
     \href{https://orcid.org/0000-0001-7151-8243}{\includegraphics[width=8pt]{orcid}}
     \textbf{Inez Maria Zwetsloot,}$^{\textrm{a,c},1,\star}$
     \href{https://orcid.org/0000-0002-6144-4188}{\includegraphics[width=8pt]{orcid}}
     \textbf{Hailiang Wang,}$^{\textrm{c}}$\textbf{ and}
     \textbf{Kwok Leung Tsui}$^{\textrm{c}}$
     \href{https://orcid.org/0000-0002-0558-2279}{\includegraphics[width=8pt]{orcid}}\\
   \end{large}
\begin{footnotesize}
  \vspace{2mm}
  $^{\textrm{a}}$Department of System Engineering and Engineering Management, City University of Hong Kong,
  Tat Chee Avenue, Kowloon, Hong Kong\\
  $^{\textrm{b}}$Department of Mathematics and Statistics, Helmut Schmidt University, Hamburg, Germany\\
  $^{\textrm{c}}$School of Data Science, City University of Hong Kong, Tat Chee Avenue, Kowloon, Hong Kong
  \vspace{2mm}\\
  $^{\star}$Corresponding author:
  \href{mailto:i.m.zwetsloot@cityu.edu.hk}{i.m.zwetsloot@cityu.edu.hk}\\
  $^{1}$Authors contributed equally
\end{footnotesize}
}

\date{}

\begin{document}	
\maketitle

\begin{abstract}
  \noindent\textbf{Background:}
  All-in-one station-based health monitoring devices are implemented in elder
  homes in Hong Kong to support the monitoring of vital signs of the elderly.
  During a pilot study, it was discovered that the systolic blood pressure was
  incorrectly measured during multiple weeks. A real-time solution was needed
  to identify future data quality issues as soon as possible.\\

  \noindent\textbf{Methods:}
  Control charts are an effective tool for real-time monitoring and signaling
  issues (changes) in data. In this study, as in other healthcare applications,
  many observations are missing. Few methods are available for monitoring data
  with missing observations. A data quality monitoring method is developed to
  signal issues with the accuracy of the collected data quickly. This method
  has the ability to deal with missing observations. A Hotelling's T-squared
  control chart is selected as the basis for our proposed method.\\

  \noindent\textbf{Findings:}
  The proposed method is retrospectively validated on a case study with a known
  measurement error in the systolic blood pressure measurements. The method is
  able to adequately detect this data quality problem. The proposed method was
  integrated into a personalized telehealth monitoring system and prospectively
  implemented in a second case study. It was found that the proposed scheme
  supports the control of data quality.\\

  \noindent\textbf{Conclusions:}
  Data quality is an important issue and control charts are useful for real-time
  monitoring of data quality. However, these charts must be adjusted to account
  for missing data that often occur in healthcare context.
\end{abstract}
\section{Background}
Applications of Telehealth are growing due to the fast development of sensor
technology. This has enabled the development of relatively cheap and easy-to-use
devices for (self-)evaluation of health indicators and well-being. These
techniques have the potential to help current elder care providers to track
vital signs, detect physiological changes and predict health risks.

\begin{figure}[!htb]
  \centering
  \includegraphics[width=0.7\textwidth]{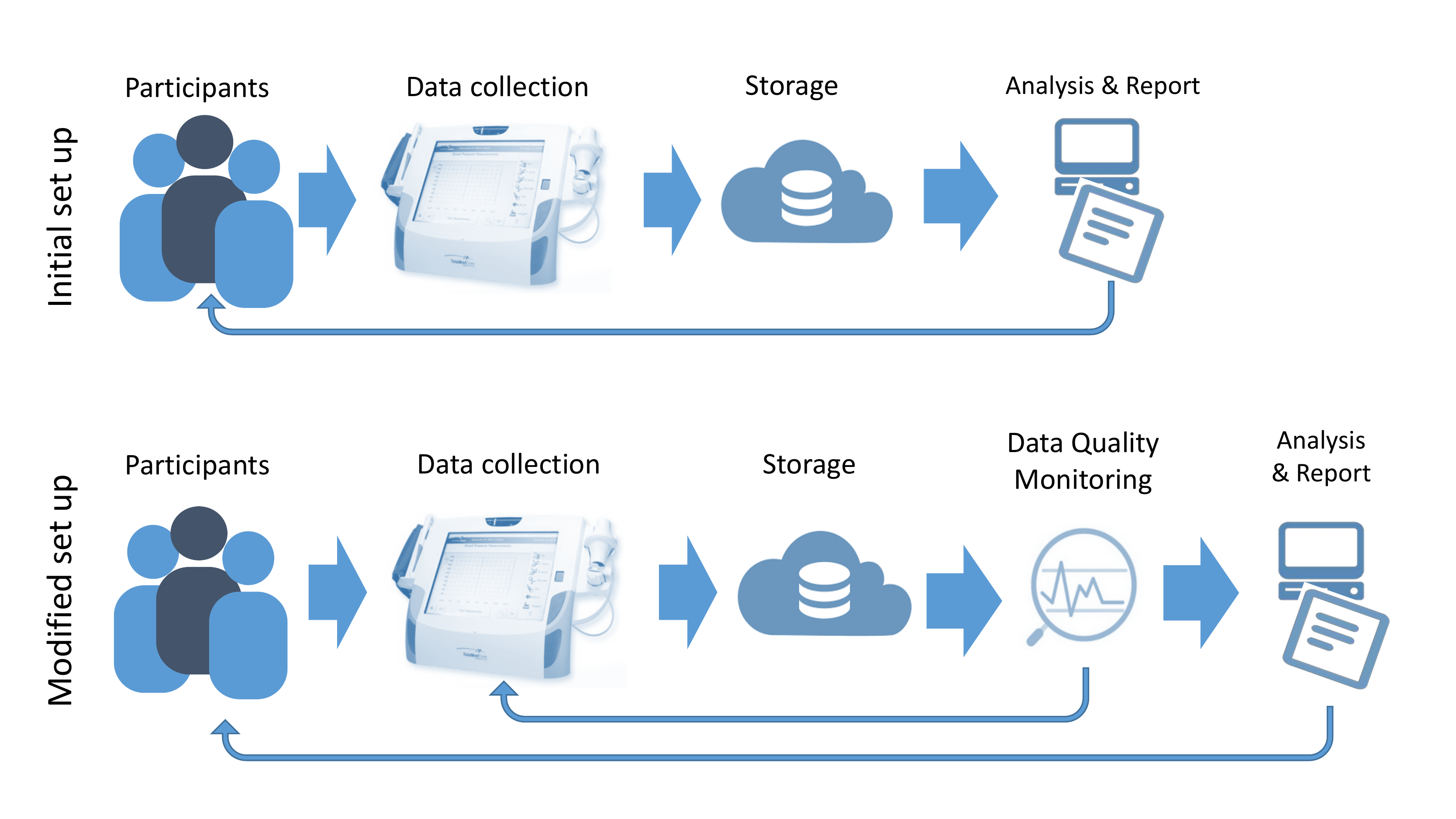}
  \caption{Overview of the telehealth system, initial (top) and modified (bottom).}
  \label{fig:datacollectionsystem}
\end{figure}

A pilot study has been conducted with an all-in-one station-based telehealth
device in Hong Kong to help track elder's vital signs. This telehealth system
is designed to provide computer-aided decision support for clinicians and
community nurses. It also enables them to easily monitor and analyze an elder's
vital signs and well-being. For more details on this system see Yu and colleagues
\cite{Yu.etal_2017}. The present study involved two elder care centers. In
each center, elder's volunteered to have their vital signs measured daily.
For approximately three months, trained and qualified research staff visited
the centers and assisted the elders to accomplish the measurements on each
measurement day (usually five days a week). Vital signs were measured using
a commercial all-in-one station-based telehealth device (TeleMedCare, Health
Monitor, TeleMedCare, Sydney, Australia). Data is stored on a server, processed,
analyzed and summarized into a report which is given to the participants.
The framework of this telehealth monitoring system is shown in the top panel
of \figref{datacollectionsystem}.

In the pilot study, we found that the systolic blood pressure was incorrectly
recorded during multiple weeks. A structural data quality method was needed
to ensure the accuracy of vital signs data collection. The proposed data quality
monitoring is implemented on a daily basis with a feedback loop to the telehealth
monitoring system, as illustrated in the lower panel of \figref{datacollectionsystem}.
\section{Methods}
\subsection{Settings}
The elder care center where the systolic blood pressure was incorrectly measured
(center A), is an elder day care center situated in Kowloon, Hong Kong. The pilot
study in center A had 24 participants and was conducted in a period from 18.12.2017
to 07.03.2018. For each participant, five vital signs were measured on each
measurement day; body temperature (BT) in degrees Celsius ($^{\circ}{\rm C}$),
systolic and diastolic blood pressure (SBP and DBP) in millimeters of mercury
(mm Hg), heart rate (HR) in number of beats per minute and peripheral capillary
oxygen saturation (SpO$_2$) in percentages (\%). We also included a second center
(B) in our pilot study. In center B there were 12 participants and the study ran
from 01.03.2018 to 31.05.2018.

\begin{figure*}[!htb]
  \centering
  \includegraphics[width=1\textwidth]{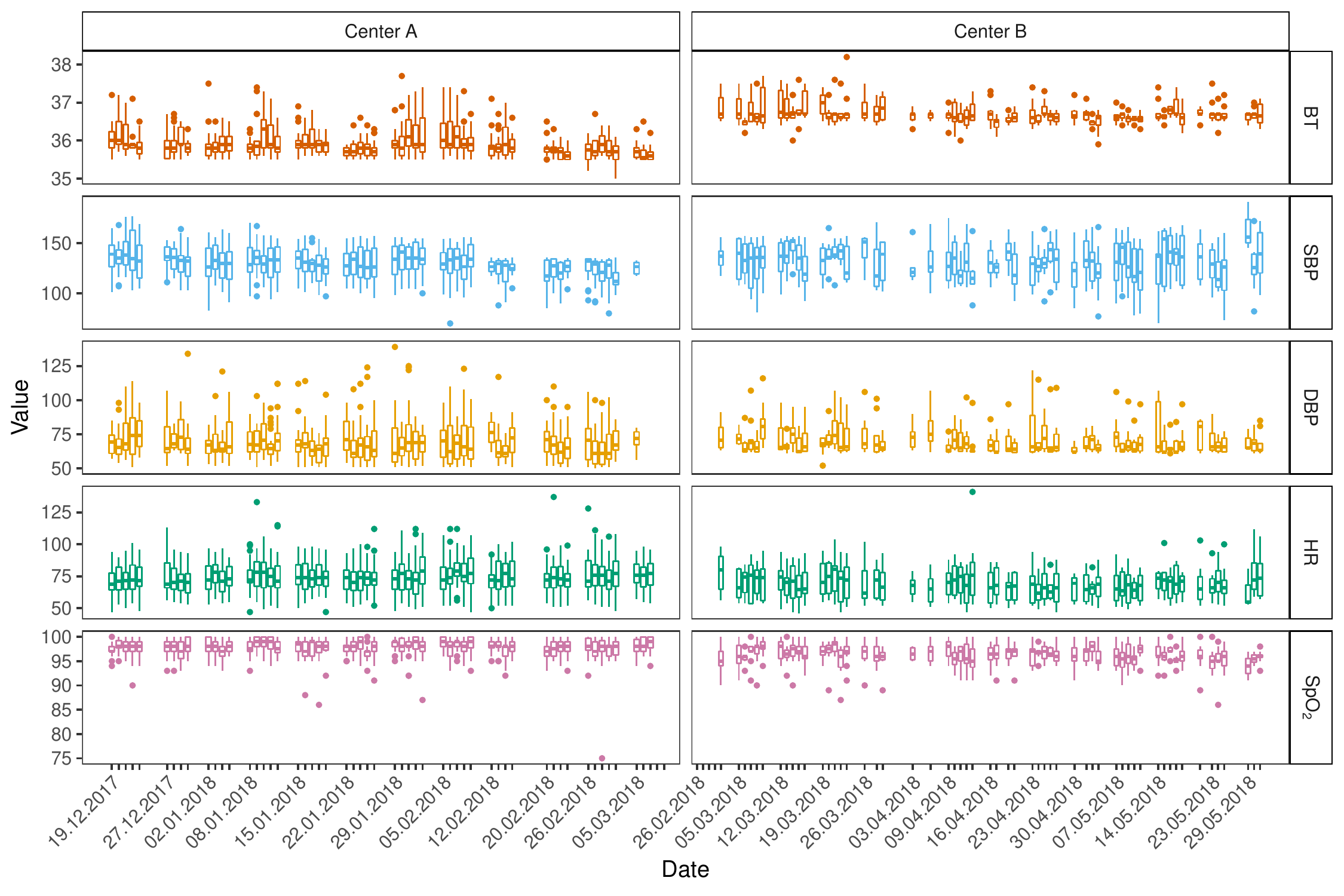}
  \caption{Boxplots of vital signs for elders in center A (left panel) and B (right panel).}
  \label{fig:boxplots}
\end{figure*}

\figref{boxplots} gives an overview of the collected data, with the panels showing
the vital signs for all participants in center A (left) and center B (right)
over the entire study period. Note that initial data is cleaned by deleting
improbable low values (i.e. outliers such as HR values of zero).

Close evaluation of \figref{boxplots} showed that starting from 12.02.2018 the
SBP measurements in Center A do not exceed 136, which is an abnormally low
maximum value for SBP measurements. Retrospective investigation revealed that
the SBP measurement subsystem was accidentally and unknowingly limited to a
maximum value of 136. Hence, the collected data was capped for the rest of the
study period. This data quality issue affected all study participants in center A.
\subsection{Choice of solution}
\label{choice}
In industrial applications, measurement system analysis (MSA) is performed to
verify the accuracy of the measurement system. MSA studies evaluate the
repeatability and reproducibility of the measurement system. And focus on the
measurement system as a whole, including the measurement device, the people
operating the device as well as the environment \cite{Montgomery_2007}. A
control chart is an associated tool often used to visualize measurement
variability. In this paper, a data quality monitoring method based on a control
chart is proposed to improve the measurement accuracy of the data collected
with the telehealth device.

A control chart is a statistical and visual tool. It is designed to prospectively
signal change(s) in data streams quickly. Control charts have been used in clinical
settings \citep{Waterhouse.etal_2010}, as well as for monitoring the quality of
cardiac surgery \citep{Gan.etal_2017} and for monitoring in epidemiologic studies
\citep{Harel.etal_2008}. Overviews of control charts in healthcare settings can be
found in  \citep{Woodall_2006, thor.etal_2007,Tennant.etal_2007}. More specifically,
single patient univariate and multivariate monitoring of vital signs is done in
\cite{Ma.Lee_2014} who used deleted cases. Corn{\'{e}}lissen and colleagues
\cite{Cornelissen.etal_1997} implement self-starting CUSUM chart for vital signs monitoring.
Sparks and colleagues \cite{Sparks.etal_2016} monitor vital sign trends of a single patient with univariate
exponential weighted moving average control charts and multiple vital signs with
dynamic Biplots.

A control chart can also be used to monitor data quality by detecting changes in
the data collection system as timely as possible. Jones-Farmer and colleagues
\cite{Jones-Farmer.etal_2014} provide a framework for control charts and data
quality monitoring. An error in the measurement system will show up as a change
in the level of the vital sign(s). Hence, a change in the level of vital sign(s)
should be signaled.

A standard method used to monitor multiple variables is the Hotelling's $T$-squared
control chart, for example see \citep{Waterhouse.etal_2010, Rigdon.Fricker_2015}.
As our all-in-one station-based telehealth device records multiple vital signs
for each participant we employed a Hotelling's $T$-squared control chart. The
choice for a multivariate, rather than multiple univariate charts is motivated
medically as vital signs are known to be correlated, for example SBP and DBP
\cite{Gavish.etal_2008}. In section \ref{sec:findings} we verify this when we implement our case
studies.

Each day we obtain data from multiple participants consecutively. In this work,
we choose to treat the data as grouped and obtain each day a data matrix containing
all vectors of each individual (see section \ref{Method} for details). We have three reasons
to adopt this approach, rather than treating the data as individual data. Firstly,
it fits with the nature of our data collection: the data collected with the telehealth
device were synchronized daily after all participants have undergone the measurements.
Secondly, subgrouping our data helps with the missing data issue. We create from the
subgrouped data a vector of averages. Hence a matrix with a lot of missing data is
converted to a vector which is nearly always complete (see section \ref{Method} for details).
Thirdly, by subgrouping and averaging our data we create approximately independent
and normally distributed data. Finally, this leads to the convenience of monitoring
a single control chart rather than one chart for each participant.
\subsection{Missing data}
\label{missingdata}
\begin{figure*}[!htb]
  \centering
  \includegraphics[width=1\textwidth]{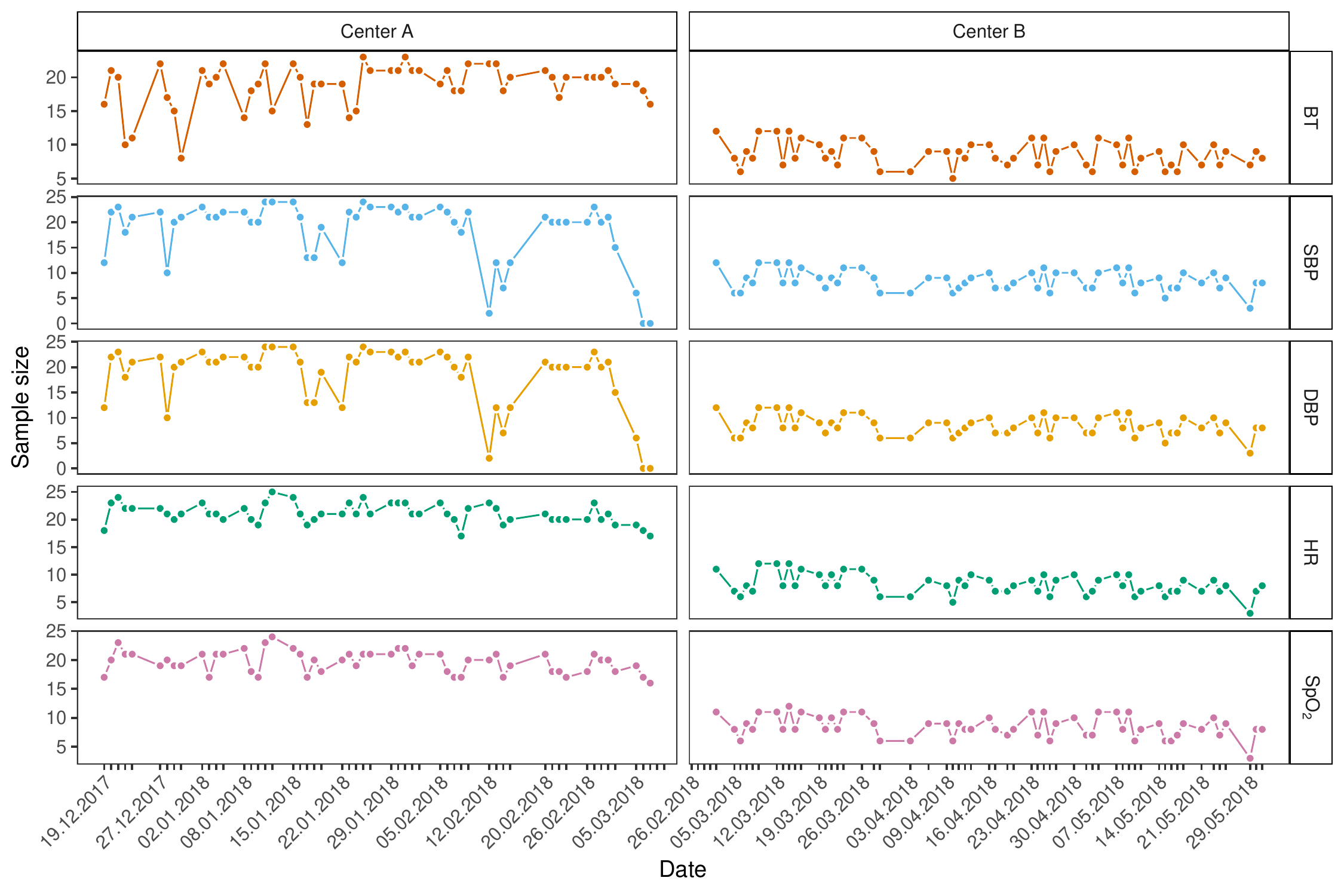}
  \caption{Sample size of vital signs for elders in center A (left panel) and B
    (right panel).}
  \label{fig:samplesizes}
\end{figure*}

In observational studies missing data are often encountered (cf.
\cite{Needham.etal_2009}) and our study is no exception. We encounter three types
of missing data:
\begin{enumerate}
  \item Not every vital sign is obtained daily.
  \item Due to the type of elder care centers (day care) not all elders show up
    everyday for measurements.
  \item Some elders join the program late or drop out early. Resulting into a long
    streams of missing data at the start or end of the study period, which is a
    common situation in healthcare related studies \citep{Celler.etal_2018}.
\end{enumerate}
\figref{samplesizes} displays the varying number of obtained measurements (elders)
over time for each vital sign and center. This varying sample size shows that we
have many missing data (without missing data all sample sizes would be equal to 24
of center A and 12 for center B). Overall 10.3 \% of the data is missing in center
A. This percentage varies across the vital signs from 1.7\% up to 14.3\%. Overall
3.9\% of the data is missing in center B. The percentage varies across the vital
signs from 2.4\% up to 6.3\%.

The literature \citep{schafer_2002} categorizes random missing data into two types:
missing at random (MAR) and missing completely at random (MCAR). The propensity of
the MAR observations is due to the random in-out of the elders and machining problems
while observations in the data set are MCAR when a random event (e.g., typhoon)
occurred and it is independent of the machining problems and presence of elders.
From the available information, it is assumed that the data is missing at random
(MAR) in our study.

Traditional control charts are designed for complete data sets and it is difficult
to run charts with missing data. A solution to apply control charts to our incomplete
dataset (addressed above as first type of missing data) with varying number of elders
on each time point (afore-mentioned missing data type two and three) is needed.
One way to deal with missing data, in a healthcare setting, is to perform data
imputation to ``fill in'' the missing values. This approach is taken by
\citep{Waterhouse.etal_2010} for the multivariate monitoring in a clinical setting
and by \citep{Herbert.etal_2011} who use imputation in a longitudinal study. The
effect of using different imputation methods on the performance of multivariate
control charts is studied by \citep{Madbuly.etal_2013} and \citep{Mahmoud_2014}.
Both studies concluded that control charts based on imputation methods provide
better performance as compared to control charts based on deletion method.
Currently, imputation is the selected method when missing observations are
encountered. However, \citep{Penny_2012} showed that imputation can only work
properly when the percentage of missing data is small and when we know this
percentage a-priori. For our settings, data imputation is complicated because
of the afore-mentioned third type of missing data: late joining and dropping
of elders. This is uncontrollable and the number of late joiners is unknown
a-priori. Therefore, it is unknown a-priori what the percentage of missing data
will be.

Another way to deal with missing data is to use control charts for variable
sample sizes (i.e. varying number of participants per day). Previous studies
proposed some methods for univariate data \citep{Huang.etal_2016a, Aslam.etal_2016}
and some for multivariate data \citep{aparisi.etal_1996, Kim.Reynolds_2005}.
The latter methods assume that the varying sample sizes are known a-priori
and that the researcher controls the number of samples. Hence these methods
are not straightforward applicable here.

To the best of our knowledge, none of the existing control charts for variable
sample sizes has the ability to adequately deal with random varying sample
sizes in a multivariate setting. Hence, a new method is developed by adapting
the Hotelling's $T$-squared control chart to accommodate for missing data and
random varying sample size without the need to impute missing data.
\subsection{Method development}
\label{Method}
For the purpose of method development, each participant is indexed by
$k \left( k = 1,2, \ldots ,\;n \right)$, each vital sign by
$j \left( j = 1,2, \ldots ,\;p \right)$ and each day by $i$. Let $X_{ijk}$
be the measurement of person $k$ for vital sign $j$ on day $i$. Hence, $n$
vectors  $\mathbf{X}_{ik} \in \mathbb{R}^p$ are obtained each day. Let
$\boldsymbol{\mu}$ denote the expectation and $\boldsymbol{\Sigma}$ the
covariance matrix of $\mathbf{X}_{ik}$

To monitor the data quality, the average value for each vital sign is computed
by averaging over those patients who showed up for measurement on day $i$.
Giving vector $\mathbf{\bar{X}}_i$ of $p$ elements $(\bar{X}_{i1},\dots,\bar{X}_{ip})^{\top}$.
Here, $\bar{X}_{ij}$ is the average of $n_{ij}\leq n$ elders for vital sign
$j$. The number of elders $n_{ij}$ depends on the day $i$ and the vital sign
$j$, and the changing $n_{ij}$ allow us to model the missing data ($n_{ij}$).
In Table \ref{tab:datastructure} we show the data structure, in this example elder
\#2 did not show up and therefore its data is missing as indicated by a star.
Also the first vital sign for elder \#1 was not recorded. The right column shows
the corresponding sample sizes for each vital sign in this example, the actual
values of $n_{ij}$ are displayed in \figref{samplesizes}.

\begin{table}[h]
  \centering
  \caption{Data structure, missing data indicated by a *.}
  \begin{tabular}{*{11}{c}}
    $\mathbf{X}_{i1}$
    & $\mathbf{X}_{i2}$
    & $\mathbf{X}_{i3}$
    & $\cdots$
    & $\mathbf{X}_{ik}$
    & $\cdots$
    & $\mathbf{X}_{in}$
    &
    & $\mathbf{\bar{X}}_{i}$
    &
    & $\mathbf{n}_{i}$\\
    \cmidrule[.7pt]{1-7}
    \cmidrule[.7pt]{9-9}
    \cmidrule[.7pt]{11-11}
    *
    & *
    & $X_{i13}$
    & $\cdots$
    & $X_{i1k}$
    & $\cdots$
    & $X_{i1n}$
    & $\rightarrow$
    & $\bar{X}_{i1}$
    &
    & $n_{i1} $\\
    $X_{i21}$
    & *
    & $X_{i23}$
    & $\cdots$
    & $X_{i2k}$
    & $\cdots$
    & $X_{i2n}$
    & $\rightarrow$
    & $\bar{X}_{i2}$
    &
    & $n_{i2} $\\
    $\vdots$
    & $\vdots$
    & $\vdots$
    & $\vdots$
    & $\vdots$
    & $\vdots$
    & $\vdots$
    & $\vdots$
    & $\vdots$
    &
    & $\vdots$ \\
    $X_{ip1}$
    & *
    & $X_{ip3}$
    & $\cdots$
    & $X_{ipk}$
    & $\cdots$
    & $X_{ipn}$
    & $\rightarrow$
    & $\bar{X}_{ip}$
    &
    & $n_{ip}$\\
    \cmidrule[.7pt]{1-7}
    \cmidrule[.7pt]{9-9}
    \cmidrule[.7pt]{11-11}
  \end{tabular}
  \label{tab:datastructure}
\end{table}

In section \ref{missingdata}, we argued that our data is MAR. Under this MAR
assumption it follows that  $\boldsymbol{\mu}_{\mathbf{\bar{X}}}=\boldsymbol{\mu}$
and after some derivation (see \hyperref[app:derivation]{Appendix}), we also have
\begin{equation} \label{eq:sigma-xbar}
  \boldsymbol{\Sigma}_{\mathbf{\bar{X}}_i} = \mathbf{W}_i \odot \boldsymbol{\Sigma}.
\end{equation}
Here, $\odot$ denotes element wise matrix multiplication, which is also known
as Hadamard multiplication and $\mathbf{W}_i$ is a matrix weighting the elements
of $\boldsymbol{\Sigma}$ according to the number of paired observations
(participants) available for measurement on day $i$. The matrix $\mathbf{W}_i$
is a $p\times p$ matrix defined as
\begin{equation} \label{eq:vee}
  \mathbf{W}_i = \left[ \frac{ \left| U_{ij} \cap U_{ij'} \right| }{n_{ij}n_{ij'}} \right]_{j,j'=1,\dots,p}.
\end{equation}
Here, $U_{ij}$ is the set of participants for whom a measurement of vital sign $j$
on day $i$ is available and $n_{ij}=|U_{ij}|$ is the number of elements in $U_{ij}$,
i.e. the number of participants. For example, if on day $i$ two participants, say
$\#1$ and $\#3$, showed up to measure vital sign $j=2$ then $U_{i2}=\{1, 3\}$ and
$n_{i2}=2$. Note that under complete observations, it follows that $n_{ij}=n$ and
hence $\mathbf{W}_i = [1/n]$. It is possible that $n_{ij} = 0$ when one of the vital
signs $j$ is not measured for all participants on the day $i$. This happened on the
final two days in center A for SBP and DBP. To account for this we reduce the
dimension of our monitoring statistic. This will be discussed in detail below.

The design of a Hotelling's $T$-squared monitoring statistic is
$T^2_i = (\mathbf{\bar{X}}_i-{\boldsymbol{\mu}})^{\top}
\boldsymbol{\Sigma}_{\mathbf{\bar{X}}_i}^{-1} ( \mathbf{\bar{X}}_i-{\boldsymbol{\mu}})$.
In order to compute this statistic, estimates of the
unknown parameters $\boldsymbol{\mu}$ and $\boldsymbol{\Sigma}_{\mathbf{\bar{X}}_i}$
are needed. For this we select the first 19 days from each study for estimation
(as the Phase I data set). To avoid biased estimates due to outliers in the data
(see \figref{boxplots}), estimates of the mean vector $\boldsymbol{\hat{\mu}}$ and
the covariance matrix $\boldsymbol{\hat{\Sigma}}$ are obtained using a robust
estimation method. Various robust methods have been evaluated \citep{Alfaro.Ortega_2009}
for the Hotelling's $T$-squared control chart.
Here, the orthogonalized Gnanadesikan–Kettenring (OGK) estimation method
\citep{Maronna.Zamar_2002} is used, because it provides positive definite and
approximately affine equivariant robust estimates. The OGK estimators are obtained
using the R-package \textsf{rrcov} \citep{Todorov.Filzmoser_2009}. Recall that
$\boldsymbol{\hat{\Sigma}}_{\mathbf{\bar{X}}_i} = \mathbf{W}_i \odot \boldsymbol{\hat{\Sigma}} $
by using the OGK estimator we obtain $\boldsymbol{\hat{\Sigma}}$ based on the
Phase I data which we clubbed and omitted incomplete cases. We argued that
the data are missing MAR, therefore this should yield a unbiased and consistent
estimator of $\boldsymbol{\Sigma}$ and hence also of $\boldsymbol{\Sigma}_{\mathbf{\bar{X}}_i}$.

The final monitoring statistic now becomes:
\begin{equation} \label{eq:hotelling}
  T^2_i = (\mathbf{\bar{X}}_i-{\boldsymbol{\hat{\mu}}})^{\top} (\mathbf{W}_i \odot
  \boldsymbol{\hat{\Sigma}})^{-1} ( \mathbf{\bar{X}}_i-{\boldsymbol{\hat{\mu}}})
\end{equation}
and the control chart signals, when $T^2_i$ exceeds the Upper Control Limit (UCL).
Whenever $T^2_i$ exceeds the UCL, a signal is observed. This signal
should be investigated and appropriate corrective action should be taken. Usually,
for multivariate control charts, a decomposition method is used to determine which
variables are responsible for the signal. For our proposed Hotelling's $T$-squared
control chart, the Mason-Young-Tracy (MYT) decomposition method \citep{mason.etal.1995}
was adopted. In the MYT decomposition, $T^2$ statistics are calculated for all possible
subsets of vital signs and plotted against the respective UCL to identify the vital sign
or combination of vital signs, which give a signal.

The UCLs for all possible combinations of vital signs are obtained by using the
following simulation procedure:

\begin{enumerate}
\item Generate a data set of $m \times \bar{n}$ vectors
  $\mathbf{X}_i \sim \mathcal{N}(\boldsymbol{\hat{\mu}}, \boldsymbol{\hat{\Sigma}})$.
  Here $m$ represents the number of days of data used to estimate the mean and covariance.
  And $\bar n = \frac{1}{mp} \sum\limits_{i=1}^m  \sum\limits_{j=1}^p n_{ij}$ is
  defined as the overall average number measurements for the vital signs obtained
  on a day.
\item Based on this data set, compute the robust mean vector
  $\boldsymbol{\hat{\mu}}_{\mathrm{OGK}}$ and covariance matrix
  $\boldsymbol{\hat{\Sigma}}_{\mathrm{OGK}}$ by using the OGK estimation method.
\item Generate vector $\mathbf{\bar{X}}_i \sim \mathcal{N}(\boldsymbol{\hat{\mu}},
  \boldsymbol{\hat{\Sigma}}/\bar{n})$. Compute the Hotelling's $T$-squared statistic as
  $T^2_i = (\mathbf{\bar{X}}_i-{\boldsymbol{\hat{\mu}}_{\mathrm{OGK}}})^{\top}
  \boldsymbol{\hat{\Sigma}}_{\mathrm{OGK}}^{-1} ( \mathbf{\bar{X}}_i-{\boldsymbol{\hat{\mu}}_{\mathrm{OGK}}})$.
\item Repeat step 3 for 10,000 times and select the $(1-\alpha)$-th quantile of
  $T^2$ as UCL.
\end{enumerate}
Repeat steps 1-4, 100 times and calculate the final UCL as the average of all
obtained control limits from step 4. For both center A and B, $m$ equals $19$.
For center A $\bar{n}$ equals $20$ and for center B $\bar{n}$ equals $9$. From
the simulations, we get $\mathrm{UCL}=17.31$ for center A and $\mathrm{UCL}=18.59$
for the center B. These UCLs are obtained by fixing the false alarm rate $\alpha=0.02$,
which implies a false alarm every 50 days on average. To obtain these UCLs, we have
assumed complete data in our simulation procedure. This is a simplification to
facilitate easy simulation. We have also implemented the simulation procedure with
varying missing data scenarios. The obtained UCLs for these scenarios are all
comparable to the limits discussed above and they differ maximally at a level of 5\%.

One additional modification to the chart (reduction in vital signs) is necessary.
Whenever a vital sign is not measured for at least one participant, one element
in $\mathbf{\bar{X}}_i$ will be missing. The corresponding row from
$(\mathbf{\bar{X}}_i-\boldsymbol{\mu})$ is then removed, the corresponding row and
column from $\boldsymbol{\Sigma}_{\mathbf{\bar{X}}_i}$ are also removed and finally,
the control limit is adjusted for that time instance.
For example, in center A, only three vital signs are measured on the last two days
of the study (see \figref{samplesizes}), so a reduced UCL is plotted at $13.29$.

Finally, before this chart can be implemented, it is important to verify that the
data comply with the important underlying assumptions of our method. In this study,
it is assumed that the average vector $\mathbf{\bar{X}}_i$ follows a multivariate
independent normal distribution with a mean vector $\boldsymbol{\mu}_{\mathbf{\bar{X}}_i}$
and covariance matrix $\boldsymbol{\Sigma}_{\mathbf{\bar{X}}_i}$. Note that we
assume normality for the average, this is rather convenient, as vital signs will
not be normally distributed on an individual basis. However for mean vectors we
can expect that normality is a reasonable assumption. We verify this assumption
by performing several multivariate normality tests; Mardia's test \citep{Mardia_1970}
and Henze-Zirkler's multivariate normality test \citep{Henze.Zirkler_1990}
implemented in the R-package \textsf{MVN} \citep{Selcuk.etal_2014}. Next we need
to consider independence. It is known that vital signs show autocorrelation
\cite{Hryniewicz.Kaczmarek-Majer_2018}. For our application we consider the
average vector, any autocorrelation in the individual vital signs will decline by
averaging over observations. We verify that the average vector is approximately
independent over time using the standard \textsf{acf} function in R. A third
important step we take is the estimation of our $\boldsymbol{\mu}$ and $\boldsymbol{\Sigma}$ matrix. We
use the robust OGK estimator using clubbed complete cases data for the first 19
days. Here we assume that this estimator is unbiased and consistent even though
we use a subset of data. It is difficult to check this empirically, however the
authors \citep{Maronna.Zamar_2002} of the OGK estimator state in their paper that this
estimator is robust to various types of outliers. Ideally we would prefer an Phase
I estimator that is also shown to be robust to missing data and deviations from
normality. However, to our knowledge an estimator for the variance-covariance matrix
that can handle outliers, missing data as well as slight deviations from normality
in sub grouped data does not (yet) exist. We welcome new research on this issue.
We compared all existing estimations methods and from both a empirical as well as
theoretical point of view we believe the OGK estimator served our purpose best.
\section{Findings}
\label{sec:findings}
Our control chart is designed to detect data quality issues. A signal may show that
the SBP measurement sub-system is capped (the reason for this project). However, a
signal can also be caused by changes in individual participant's vital signs. This
signal than may point towards the need for medical assistance (rather than measurement
system calibration). In the elder care centers, well trained personal takes care of
the elders. In this project, we consulted with them after any signal caused by
individual vital sign levels. A separate project is being conducted to design a
monitoring system for individuals. However, before we can monitor individual's vital
signs for changes in health, we need to ensure that the collected data is accurate.
That is where our proposed method is applied. The proposed monitoring method is first
validated on center A and next prospectively run on center B to monitor the data quality.
\subsection{Retrospective implementation for center A}
\label{caseA}
Next the proposed method is retrospectively implemented for center A.
The first 19 days, from 19.12.2017 to 16.01.2018, are used to obtain the OGK estimates:
\begin{align*}
  \boldsymbol{\hat{\mu}} = \left(
  \begin{array}{*{20}{l}}
    \hat{\mu}_{\mathrm{BT}}  \\
    \hat{\mu}_{\mathrm{SBP}} \\
    \hat{\mu}_{\mathrm{DBP}} \\
    \hat{\mu}_{\mathrm{HR}}  \\
    \hat{\mu}_{\mathrm{Sp{O_2}}}
  \end{array} \right) = \left(
  \begin{array}{*{20}{r}}
    35.93 \\
    131.31 \\
    67.55 \\
    73.38 \\
    97.94
  \end{array} \right)
\end{align*}
and
\begin{align*}
  \boldsymbol{\hat{\Sigma}} &= \begin{bmatrix}
    \addtolength{\tabcolsep}{-0.1cm}
    \begin{tabular}{rrrrr}
      0.10 & -0.01 & -0.06 & 0.28 & 0.00 \\
      -0.01 & 254.92 & 22.33 & -48.58 & 0.56 \\
      -0.06 & 22.33 & 87.13 & 3.54 & -0.04 \\
      0.28 & -48.58 & 3.54 & 130.38 & 5.18 \\
      0.00 & 0.56 & -0.04 & 5.18 & 2.36
    \end{tabular}
  \end{bmatrix}.
\end{align*}
Before we plot the proposed control chart, we validate the most important
assumptions of normality and independence over time. We also study the correlation
between the vital signs to motivate the multivariate nature of our method. We
found no evidence of deviation from multivariate normality. In addition we
compute the autocorrelation function and found no evidence of significant
autocorrelation between the mean levels of each of the vital signs. Finally,
we looked at the correlation matrix, standardized from the variance-covariance
matrix above, and concluded that multiple pairs of vital signs are correlated as
validated by Bartlett's test of sphericity \cite{Bartlett_1951} (p-value $< 0.0001$).

\begin{figure}[!htb]
	\centering \includegraphics[width=.75\textwidth]{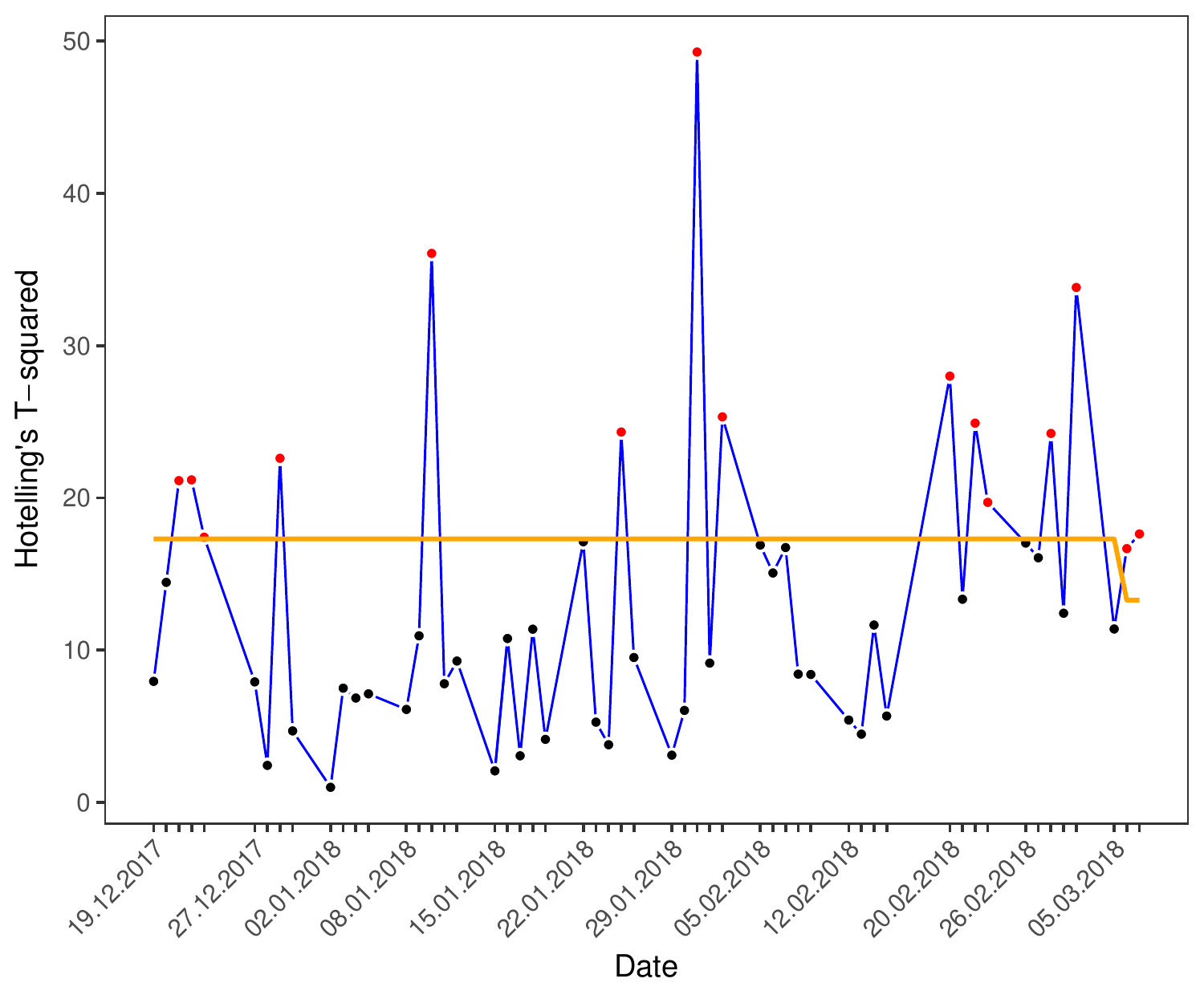}
	\caption{Hotelling's $T$-squared chart for center A.}
	\label{fig:hotellingA}
\end{figure}

Next, the control chart is plotted to monitor the data quality in case A. The
Hotelling's $T$-squared statistics are calculated by using Equation \eqref{eq:hotelling}
and plotted against the UCL=17.31 in \figref{hotellingA}.

The following catches the eye:
\begin{itemize}
\item The first four signals are observed in the interval from 21.12.2017 to
  29.12.2017. The MYT decomposition method identifies all signals as caused by
  changes in the mean of the DBP measurements.
\item The next four signals are observed in the interval from 10.01.2018 to
  02.02.2018. MYT decomposition method classifies the $5th$, and $7th$
  signal as driven by variation in the DBP and BT measurements. The $6th$
  signal, on 25.01.2018, is due to the variation in $\mathrm{SpO_{2}}$ measurements.
  The $8th$ signal, on 02.02.2018, is due to variation in $\mathrm{DBP}$.
\item The signals from 20.02.2018 until 02.03.2018 were expected as these signals
  are due to the capped measurements of SBP. However, the signal is delayed by
  four days. The MYT decomposition indicates that apart from SBP the BT also
  influenced the signal.
\item In the last two days, two signals are observed. The MYT decomposition
  indicated a decrease in the means of the BT measurements.
\end{itemize}

Overall, the collected data are not stable in this pilot study. Many issues,
especially with the SBP measurement are discovered. Hence, it is necessary to
implement a data quality monitoring method into the telehealth system
(\figref{datacollectionsystem}). Such a method can help to detect undesirable
situation as soon as possible so that appropriate action can be taken. Apart
from the monitoring scheme, other actions such as training of staff and standardized
working procedures were also applied to ensure repeatable and reproducible measurements.
\subsection{Prospective implementation for center B}
\label{caseB}
To evaluate the usefulness of our developed Hotelling's $T$-squared control chart,
we prospectively implement the developed control chart for center B. Thereby, we
verify in real-time, whether the telehealth system is functioning normally. The
data collection started on 02.03.2018 and the first week of observations are used
to estimate the parameters ($\boldsymbol{\hat{\mu}}$ and $\boldsymbol{\hat{\Sigma}}$).
A signal was observed on 09.03.2018, after careful investigation and MYT decomposition,
it was discovered that the DBP measurements were set to a maximum level on that day.
This issue was solved by adjusting the telehealth device.

\begin{figure}[!htb]
	\centering
	\includegraphics[width=.75\textwidth]{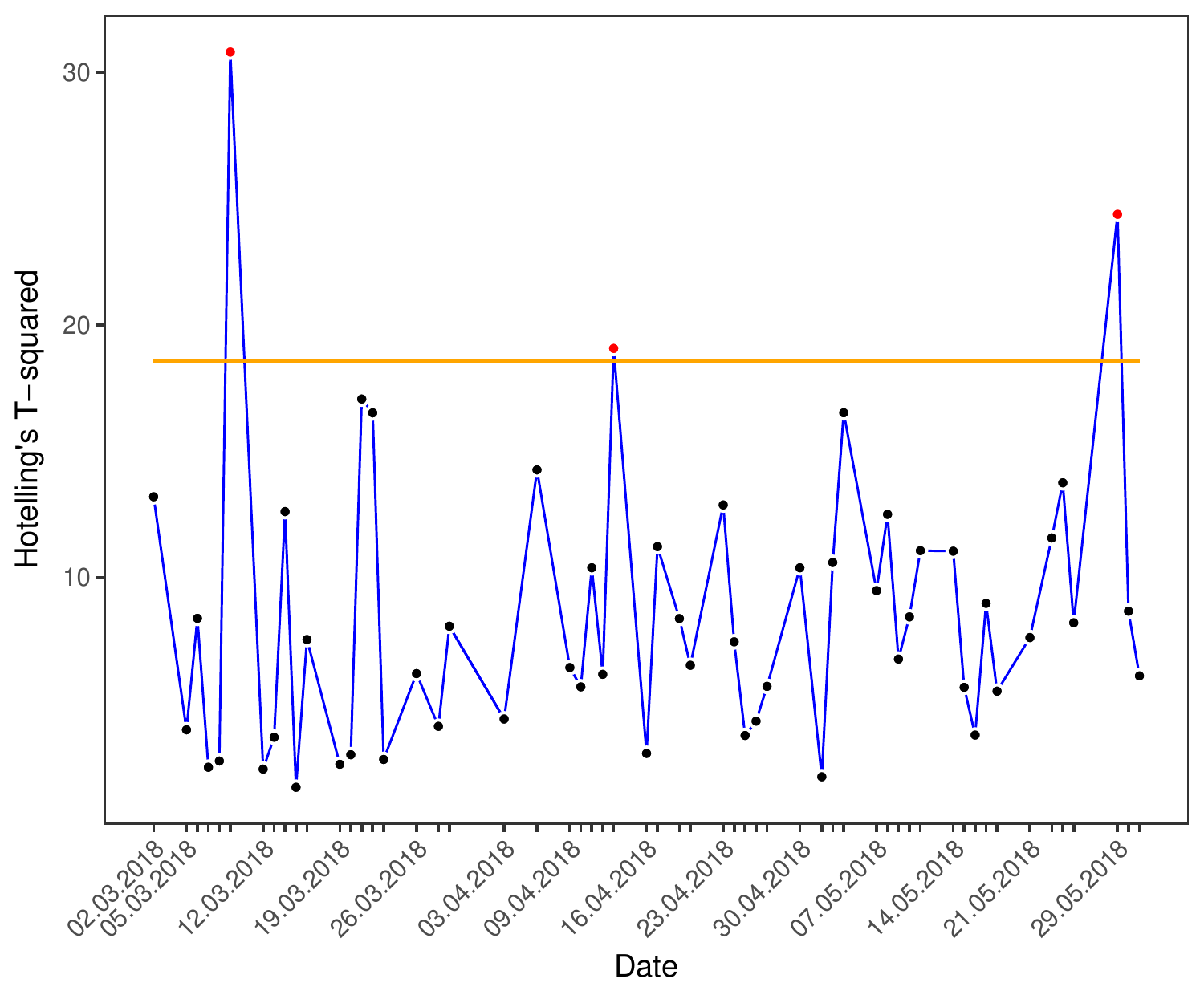}
	\caption{Hotelling's $T$-squared chart for center B.}
	\label{fig:hotellingB}
\end{figure}

The data collection was continued in the subsequent weeks and the Hotelling's
$T$-squared control chart, based on re-estimated parameters was plotted. This
iterative method was continued until $m=19$ days of data were collected. The
corresponding OGK estimates for the mean vector and covariance matrix are
\begin{align*}
  \boldsymbol{\hat{\mu}} = \left(
  \begin{array}{*{20}{l}}
    \hat{\mu}_{\mathrm{BT}}  \\
    \hat{\mu}_{\mathrm{SBP}} \\
    \hat{\mu}_{\mathrm{DBP}} \\
    \hat{\mu}_{\mathrm{HR}}  \\
    \hat{\mu}_{\mathrm{Sp{O_2}}}
  \end{array} \right) = \left(
  \begin{array}{*{20}{r}}
    36.83 \\
    133.96 \\
    69.80 \\
    71.11 \\
    96.96
  \end{array} \right)
\end{align*}
and
\begin{align*}
  \boldsymbol{\hat{\Sigma}} &=  \begin{bmatrix}
    \addtolength{\tabcolsep}{-0.1cm}
    \begin{tabular}{rrrrr}
      0.12 & 0.90 & -0.06 & 0.55 & 0.23 \\
      0.90 & 328.42 & 29.28 & 60.74 & 5.59 \\
      -0.06 & 29.28 & 69.99 & 27.13 & -0.39 \\
      0.55 & 60.74 & 27.13 & 184.52 & 0.29 \\
      0.23 & 5.59 & -0.39 & 0.29 & 3.34 \\
    \end{tabular}
  \end{bmatrix}.
\end{align*}
Similarly, as in Case A, we have validated assumptions; we found no statistical
significant difference from normality nor any evidence for autocorrelation.
Furthermore, some pairs of vital signs show correlation (Bartlett's test
p-value $< 0.0001$). The Hotelling's $T$-squared control chart for the full study
period is displayed in \figref{hotellingB}.

A second signal was observed on 13.04.2018, after careful investigation the signal
can be attributed to a very high HR for a single participant. Another signal was
observed on 29.05.2018, the MYT decomposition method identified that the cause for
the signal is an unusual high SBP level and an unusual low $\mathrm{Sp{O_2}}$ level.

Overall, the implementation of the data quality monitoring method provides a timely
indication of data quality issues which may leads to adjustment of the telehealth
system and a decrease in the wastage of resources.

\section{Discussion and conclusion}
Telehealth applications provide many opportunities for innovations in healthcare.
However, the accuracy and reliability of measured data may be challenging to
guarantee when various people, especially non-medical experts, perform the measurements
\cite{Brewster.etal_2013, Taylor.etal_2014, Celler.Sparks_2015a}. There are many ways
to validate and guarantee data accuracy, such as training people and calibration of
the measurement system. In this article, we set forth an additional check to verify
the real-time quality of measured data. A Hotelling's $T$-squared control chart is
designed, which is modified to deal with missing data. After testing the proposed
method on a case study with known data quality issues, the control chart is prospectively
implemented on a second case study. In the second study, a data quality issue is detected
after one week, which was timely solved. Hence, a regular focus on the data quality helps
to ensure the validity and accuracy of the collected measurements and a quick feedback to
the data quality monitoring system is essential to solve problems on-time.

Data monitoring is also important on an individual level and more comprehensive models
are needed to deal with the heterogeneity of individuals. An individual monitoring tool
can be helpful to detect health changes for each elder separately. However, before this
can be done, the data has to be accurate. The developed method for data quality monitoring
can thus be seen as a first step before monitoring the health individually. The focus is
on handling the variable sample sizes, in our application due to MAR type missing data.
Our approach can also be applied to other scenarios with missing data where monitoring is
required. It would be useful if future research would compare the proposed method to
existing methods for monitoring data with missing observations, such as the imputation
method.

EWMA or CUSUM control charts are generally quicker in detection of a failure mode than
a Hotelling's $T$-squared control chart. Therefore, a possible extension for future
research could be a multivariate version of the CUSUM procedure with the dynamic
probability control limits developed by \cite{Huang.etal_2016a}.
\section*{Author contributions}
All authors made substantial contributions to the study design and methods. HW, KLT
and IZ extracted data. PW, TM and IZ analyzed and visualized the data and derived
the method. Further, all authors interpreted results and drafted the manuscript.
\section*{Funding}
This work was partly supported by the Research Grants Council Theme-Based Research
Scheme [T32-102-14N] and City University of Hong Kong [9610406, 7200548].
\section*{Acknowledgment}
We thank all participants of the biweekly elder care project meetings for helpful
comments and feedback. We thank all reviewers for the constructive comments which
have helped us to substantially improve and clarify the manuscript.
\section*{Conflicts of interest}
The authors declare no conflict of interest.

\section*{Appendix}
\label{app:derivation}
In this appendix, we provide the derivation of the adjustment factor $\mathbf{W}_i$ as used in equation \eqref{eq:sigma-xbar}: $\boldsymbol{\Sigma}_{\mathbf{\bar{X}}_i} = \mathbf{W}_i \odot \boldsymbol{\Sigma}$.
As stated in equation \eqref{eq:vee}, $\mathbf{W}_i$ is equal to
\begin{equation*} \label{eq:vee1}
  \mathbf{W}_i = \left[ \frac{ \left| U_{ij} \cap U_{ij'} \right| }{ n_{ij} n_{ij'} } \right]_{j,j'=1,\dots,p}.
\end{equation*}
As discussed in section \ref{Method}, $U_{ij}$ is the set of elders who showed up for measureing their vital sign $j$ on day $i$ and define $n_{ij}=|U_{ij}|$.
Recall that the matrix $\boldsymbol{\Sigma}_{\mathbf{\bar{X}}_i}$ is the covariance matrix for the random mean vector $\mathbf{\bar{X}}_i$ which is defined as
\begin{equation*}
  \mathbf{\bar{X}}_i = \left[ \frac{1}{n_{ij}} \sum_{ k \in U_{ij} } X_{ijk} \right]_{j=1,\dots,p}.
\end{equation*}
Following a similar set-up as \cite{Kim.Reynolds_2005}, we derive the covariance between any two elements $j$ and $j'$ of $\mathbf{\bar{X}}_{i}$ as
\begin{align*}
  Cov  \left( \bar{X}_{ij}, \bar{X}_{ij'} \right)
      &= Cov \left(\frac{1}{n_{ij}} \sum \limits_{k \in  U_{ij}} X_{ijk}, \frac{1}{n_{ij'}} \sum \limits_{k' \in U_{ij'}} X_{ij'k'} \right) \\
      &= \frac{1}{ n_{ij} n_{ij'}} \sum\limits_{k \in  U_{ij}} \mathop \sum \limits_{k' \in  U_{ij'}}
        Cov \left( X_{ijk}, X_{ij'k'} \right).
\end{align*}
We assume independence of observations between individual elder, i.e.,
$Cov \left( X_{ijk}, X_{i{j'k'}} \right) = 0$ whenever $k \ne k'$.
Now it follows that
\begin{align*}
  Cov  \left( \bar{X}_{ij}, \bar{X}_{ij'} \right)
  &= \frac{1}{n_{ij} n_{ij'}} \mathop \sum\limits_{k \in {U_{ij}} \cap U_{ij'}} Cov \left( X_{ijk}, X_{ij'k} \right) \\
  &= \frac{ \left| U_{ij} \cap U_{ij'} \right| }{ n_{ij} n_{ij'} } Cov \left( X_{ij}, X_{ij'} \right) \\
  &= \frac{ \left| U_{ij} \cap U_{ij'} \right| }{ n_{ij} n_{ij'} } \boldsymbol{\sigma} _{jj'}.
\end{align*}
Alternatively, we can write
\begin{equation*}
  \boldsymbol{\Sigma}_{\mathbf{\bar{X}}_i} = \left[ \frac{ \left| U_{ij} \cap U_{ij'} \right| }{ n_{ij} n_{ij'} } \boldsymbol{\sigma} _{jj'} \right]_{j,j'=1,\dots,p}.
\end{equation*}
Where $\mathbf{W}_i = \big[\left| U_{ij} \cap U_{ij'} \right| / n_{ij} n_{ij'}\big]$, can be interpreted as the weighting matrix which takes the number of observed data points into account.
Hence, the covariance matrix of $\mathbf{\bar{X}}_i$ is equal to
\begin{equation*}
  \boldsymbol{\Sigma}_{\mathbf{\bar{X}}_i} = \mathbf{W}_i \odot \boldsymbol{\Sigma}
\end{equation*}
where, $\odot$ is the Hadamard product, which denotes element wise matrix multiplication and $\boldsymbol{\Sigma}$ is the covariance matrix of the individual data.
\end{document}